\documentclass[12pt]{article}
\usepackage{epsfig}

\def\gev{~{\rm GeV}}
\def\ale{\alpha_{\rm elm}}
\def\als{\alpha_{\rm s}}

\newcommand{\gsim}{\raisebox{-4pt}{$\,\stackrel{\textstyle
                                                         >}{\sim}\,$}}
\newcommand{\tr}[1]{ {\bf #1}_\perp}
\newcommand{\ov}[1]{\overline#1}

\begin{document}
\thispagestyle{empty}
\begin{flushright}
WU B 01-14 \\
PITHA 01/13 \\
NORDITA-2001-84 HE \\
hep-ph/0112274 \\
February 2002\\[5em]
\end{flushright}

\begin{center}
\end{center}
\begin{center}

{\Large\bf The handbag contribution to $\gamma \gamma \to \pi \pi$ and
$K K$} \\
\vskip 3\baselineskip

M.\ Diehl\,\footnote{Email: mdiehl@physik.rwth-aachen.de}
\\[0.5em]
{\small {\it Institut f\"ur Theoretische Physik E, RWTH Aachen, 52056
Aachen, Germany}}\\
\vskip\baselineskip

P.\ Kroll\,\footnote{Email: kroll@theorie.physik.uni-wuppertal.de}
\\[0.5em]
{\small {\it Fachbereich Physik, Universit\"at Wuppertal, 42097
Wuppertal, Germany}}\\
\vskip \baselineskip

and C.\ Vogt\,\footnote{Email: cvogt@nordita.dk}
\\[0.5em]
{\small {\it Nordita, Blegdamsvej 17, 2100 Copenhagen, Denmark}}\\
\vskip \baselineskip

\end{center}

\vskip 3\baselineskip

\begin{abstract}
We investigate the soft handbag contribution to two-photon
annihilation into pion or kaon pairs at large energy and momentum
transfer.  The amplitude is expressed as a hard $\gamma\gamma\to
q\bar{q}$ subprocess times a form factor describing the soft
transition from $q\bar{q}$ to the meson pair.  We find the calculated
angular dependence of the cross section in good agreement with data,
and extract annihilation form factors of plausible size.  A key
prediction of the handbag mechanism is that the differential cross
section is the same for charged and neutral pion pairs, in striking
contrast with what is found in the hard scattering approach.
\end{abstract}

\newpage

{\it Introduction.}  The production of pion or other hadron pairs in
two-photon collisions at high energies has long been a subject of
great interest. Recently it has been
shown~\cite{mue1994,die1998a,fre1999} that in kinematics where one of
the photons has a virtuality much larger than the squared invariant
mass $s$ of the hadron pair the transition amplitude factorizes into a
perturbatively calculable subprocess, $\gamma^* \gamma \to q \bar{q}$,
and a soft $q \bar{q} \to \pi \pi$ transition matrix element.  The
latter was termed the two-pion distribution amplitude in order to
emphasize its close connection to the single-pion distribution
amplitude introduced in the standard hard scattering
approach~\cite{bl1980}.  The two-pion distribution amplitude is the
timelike version of a generalized parton distribution, which encodes
the soft physics information in processes such as deeply
virtual~\cite{ji1996} or wide-angle~\cite{rad1998,DFJK1} Compton
scattering.

Here we are interested in the complementary kinematical region of
large $s$, large momentum transfer from the photons to the pions, and
vanishing photon virtuality.  It has long been
known~\cite{bl1981,ben1989} that for asymptotically large $s$ the
process is amenable to a leading-twist perturbative treatment, where
the transition amplitude factorizes into a hard scattering amplitude
for $\gamma\gamma \to q\bar{q}\, q\bar{q}$ and a single-pion
distribution amplitude for each pion.  This distribution amplitude is
constrained by the photon-pion transition form
factor~\cite{kro96,bro98,DKV2001}, and it has recently become
clear~\cite{vogt2000} that the perturbative contribution evaluated
with such a distribution amplitude is well below experimental data.

\begin{figure}[ht]
\begin{center}
\psfig{file=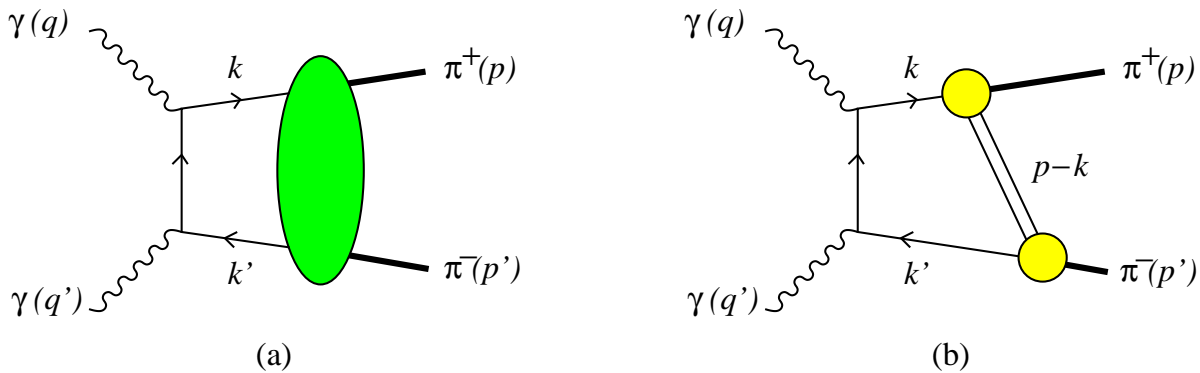,width=0.95\textwidth}
\caption{(a) Handbag factorization of the process
$\gamma\gamma\to\pi\pi$ for large $s$ and $t$. The hard scattering
subprocess is shown at leading order in $\alpha_s$, and the blob
represents the two-pion distribution amplitude. The second
contributing graph is obtained by interchanging the photon vertices.
(b) The handbag resolved into two pion-parton vertices connected by
soft partons.  There is another diagram with the $\pi^+$ and $\pi^-$
interchanged.}
\label{fig:handbag}
\end{center}
\end{figure}

In this letter we propose an approach which is complementary to the
perturbative one for large but not extremely large energies and
momentum transfers. The mechanism we investigate is similar to the one
in two-photon annihilation at large $Q^2$ but small $s$. The
corresponding diagrams have the handbag topology shown in
Fig.~\ref{fig:handbag}a, and we will express them as a hard scattering
$\gamma\gamma \to q\bar{q}$ times a form factor describing the soft
transition $q\bar{q}\to \pi\pi$ and given by a moment of the 
two-pion distribution amplitude in the kinematical region of
interest. The handbag contribution to $\gamma\gamma \to \pi\pi$
formally represents a power correction to the leading-twist
perturbative one, which will dominate at asymptotically
large scales. The approach advocated here is analogous to the handbag
contribution to wide-angle Compton scattering~\cite{rad1998,DFJK1}.

\vskip\baselineskip {\it The handbag amplitude.}  We are interested in
$\gamma\gamma$ annihilation into a meson pair at large Mandelstam
variables $s \sim -t \sim -u$.  For definiteness we consider a
$\pi^+\pi^-$ pair, the generalization to other mesons is
straightforward.  As far as possible we will proceed in analogy to the
calculation of the handbag contribution to wide-angle Compton
scattering~\cite{DFJK1}.  We wish to calculate the handbag diagrams in
the region of phase space where the $q\bar{q}\to \pi\pi$ transition is
soft.  Since the $\pi\pi$ system has large invariant mass this
requires the additional $q\bar{q}$ pair and possibly other partons
created in the hadronization process to have soft momenta.  The
momenta of the initial quark and antiquark must thus approximately
equal the respective momenta of the final state pions. We see
from Fig.~\ref{fig:handbag}b that, contrary to the crossed channel
process $\gamma\pi\to \gamma\pi$, the soft $q\bar{q}\to \pi\pi$
transition cannot be described in terms of individual pion light-cone
wave functions: the partons connecting the two pions cannot be
incoming for both of them.  We can however still understand the
diagram of Fig.~\ref{fig:handbag}b as a covariant Feynman diagram,
with each blob representing a pion-parton vertex function that is
purely soft in our kinematics.

The handbag diagrams also admit kinematical configurations where the
blob in Fig.~\ref{fig:handbag}a contains hard interactions.
Explicitly writing these as hard gluon exchange one obtains a subset
of the graphs calculated in the leading-twist perturbative approach.
Note that there are other graphs, where the two photons do not couple
to the same quark line.  At large $s$, $t$, $u$ they always require
hard gluon exchange, and thus appear in the leading-twist calculation
but not in the soft mechanism we are concerned with here.

We work in the c.m.\ frame of the reaction, with axes chosen such that
the process takes place in the 1-3 plane and the outgoing hadrons fly
along the positive or negative 1-direction.  Introducing light-cone
coordinates $v = [v^+, v^-, {\bf v}_\perp]$ with $v^\pm = (v^0 \pm
v^3) /\sqrt{2}$ for any four-vector $v$ we then have pion momenta
\begin{equation}
p = \sqrt{\frac{s}{8}}
  \left[\, 1 \, , \, 1 \, , \, \sqrt{2}\beta\, {\bf e}_1 \,\right] , 
\qquad
p'= \sqrt{\frac{s}{8}}
  \left[\, 1 \, , \, 1 \, , \, -\sqrt{2}\beta\, {\bf e}_1 \,\right] , 
\end{equation}
with the pion velocity $\beta=\sqrt{1 - 4m_\pi^2 /s}$ and ${\bf e}_1 =
(1,0)$.  We have chosen coordinate axes with the goal in mind to
describe the hadronization process by a light-cone dominated matrix
element: in our coordinate system the light-cone plus-momenta of the
hadrons appear in a symmetric way (as they do in the frame where the
handbag contribution to wide-angle Compton scattering is calculated).
The photon momenta read
\begin{eqnarray}
q &=& \sqrt{\frac{s}{8}}
      \left[\, 1+\sin\theta \, , \, 1-\sin\theta \, , 
            \, \sqrt{2} \cos\theta \, {\bf e}_1 \,\right] , 
\nonumber \\
q'&=& \sqrt{\frac{s}{8}}
      \left[\, 1-\sin\theta \, , \, 1+\sin\theta \, , 
            \, -\sqrt{2} \cos\theta \, {\bf e}_1 \,\right] ,
\end{eqnarray}
where $\theta$ is the c.m.\ scattering angle.  Up to corrections of
order $m_\pi^2 /s$ we have
\begin{equation}
 \cos\theta=\frac{t-u}{s}, 
\qquad 
 \sin\theta= \frac{2 \sqrt{t\, u}}{s} \,
\end{equation}
in terms of the usual Mandelstam variables.  In our symmetrical
reference frame the skewness, which is defined by
\begin{equation}
 \zeta=\frac{p^+}{(p+p')^+} \,,
\end{equation}
has a value of 1/2. Exploiting momentum conservation and introducing
the plus-momentum fraction
\begin{equation}
 z=\frac{k^+}{(p+p')^+} \,,
\end{equation}
we parameterize the off-mass shell quark and antiquark momenta as
\begin{equation} 
k  = \sqrt{\frac{s}{2}}
     \left[\, z, \, \bar{z} + \delta^-, 
      \, \sqrt{2z\bar{z} + \delta_\perp}\, \tr{e} \,\right] \,, 
\qquad 
k' = \sqrt{\frac{s}{2}}
     \left[\, \bar{z}, \, z - \delta^-,
      \, -\sqrt{2z\bar{z} + \delta_\perp}\, \tr{e} \,\right] \,,
\end{equation}
and their on-shell approximations as
\begin{equation} 
\tilde{k}  = \sqrt{\frac{s}{2}}
   \left[\, z, \, \bar{z}, \, \sqrt{2z\bar{z}}\, \tr{e} \,\right] , 
\qquad 
\tilde{k}' = \sqrt{\frac{s}{2}}
   \left[\, \bar{z}, \, z, \, -\sqrt{2z\bar{z}}\, \tr{e} \,\right] \,,
\end{equation}
where $\bar{z}\equiv 1-z$ and $\tr{e} = (\cos\varphi, \sin\varphi)$.
To ensure that the subprocess $q\bar{q}\to \pi\pi$ is soft we require
\begin{itemize}
\item[$(i)$] that all virtualities at the parton-hadron vertices be
soft, of order of a squared hadronic scale $\Lambda^2$,
\item[$(ii)$] and that for each parton or system of partons in a
hadron we have ${\bf k}_{\perp i}^2 /x_i^{\phantom{2}} \sim
\Lambda^2$, where $\tr{k}{}_i$ and $x_i$ respectively are the
transverse momentum and plus-momentum fraction in a frame where the
hadron moves in the positive 3-direction.\footnote{Such a frame is
obtained for each pion via a transverse boost from the c.m., see
e.g.~\protect\cite{DFJK1}.}
\end{itemize}
This enforces
\begin{equation}
 2z-1, \, \sin\varphi, \, 
 \delta^-, \, \delta_\perp \, \sim \frac{\Lambda^2}{s} \,,
\label{near-collinear}
\end{equation}
and depending on whether $\varphi\approx 0$ or $\varphi\approx\pi$
means that we have $k\approx p$ or $k'\approx p$, up to corrections of
order $\Lambda^2/s$.

We remark that condition $(i)$ alone would constrain $2z-1$ and
$\sin\varphi$ to be of order $\Lambda /\sqrt{s}$ only.  The stronger
condition $(ii)$ arises quite naturally for light-cone wave functions
\cite{bro1989}, and we also demand it here for the upper vertex in
Fig.~\ref{fig:handbag}b.  In the framework of light-cone time ordered
perturbation theory \cite{bro1989} this condition means that the
light-cone energy denominator for the intermediate state with momenta
$p$, $k-p$ and $k'$ must be soft of order $\Lambda^2$.

We now express the handbag amplitude for our process in terms of the
$\gamma\gamma \to q\bar{q}$ amplitude $H$ and a matrix element
describing the $q\bar{q} \to \pi\pi$ transition,
\begin{eqnarray}
{\cal A}=\sum_q (e e_q)^2 \int\! d^4 k \,
    \int\! \frac{d^4 x}{(2 \pi)^4} \, e^{-i\, k \cdot x} \,
    \langle \pi^+(p) \,  \pi^-(p')\, | \, T \,
    \ov{q}{}_{\alpha}(x) \, q_{\beta}(0) \, |0 \rangle \;
  H_{\alpha\beta}(k,k') \,,
\label{handbag-amp}
\end{eqnarray}
where
\begin{eqnarray}
 H_{\alpha\beta}(k,k') = \left[ \epsilon \cdot \gamma \,
    \frac{(k-q)\cdot\gamma}{(k-q)^2+i \epsilon} \, \epsilon'\cdot\gamma
  +\epsilon'\cdot\gamma \, \frac{(q-k')\cdot\gamma}{(q-k')^2+i\epsilon} \,
     \epsilon\cdot\gamma \right]_{\alpha\beta}
\label{sub-amp}
\end{eqnarray}
with the photon polarization vectors $\epsilon$ and $\epsilon'$.  The
summation index $q$ refers to the quark flavors $u$, $d$, $s$, and we
have omitted terms in $H$ suppressed by the current quark masses.

In order to select the dominant Dirac structure of the soft matrix
element, we follow \cite{DFJK1} and perform a transverse boost to a
frame where the on-shell vector $\tilde{k}'$ has a zero transverse and
hence also a zero minus-component.  In this frame we decompose the
quark field into its good and bad components,
\begin{eqnarray}
 q(0)&=&
     \frac{1}{2} \, \gamma^- \gamma^+ \, q(0)
   + \frac{1}{2} \, \gamma^+ \gamma^- \, q(0)
\nonumber \\
 &=& \frac{1}{2 k'^+} \sum_{\lambda'} \left\{
     v(\tilde{k}',\lambda') \,   
       \Big[\bar{v}(\tilde{k}',\lambda')\, \gamma^+ q(0)\Big]
   + \gamma^+  v(\tilde{k}',\lambda') \,
       \Big[\bar{v}(\tilde{k}',\lambda')\, q(0)\Big] \right\} \, ,
\label{quark-field}
\end{eqnarray}
with a sum over helicities $\lambda'/2=\pm 1/2$.  In the second line
we have used the completeness relation for massless spinors and the
relation $\tilde{k}' \cdot\gamma = k'^+ \gamma^-$ valid in the frame
we are now working in.  Since the momenta at the soft parton-hadron
vertices have large plus-components, but by definition no large
invariants, the term with the bad components is suppressed as $\Lambda
/\sqrt{s}$ compared with the good ones.  Transverse boosts leave
plus-components invariant, so that the decomposition in the second
line of (\ref{quark-field}) also holds in the c.m.\ frame.  By an
analogous argument for $\ov{q}(x)$, we obtain
\begin{eqnarray}
\lefteqn{
\ov{q}{}_\alpha(x)\, q_\beta(0) \, H_{\alpha\beta}
}
\nonumber \\
&=& \frac{1}{4 k^+ k'^+}
  \sum_{\lambda=-\lambda'} \left[
  \ov{q}(x)\, \gamma^+ u(\tilde{k},\lambda) \right] \, 
  \left[ \bar{v}(\tilde{k}',\lambda') \, \gamma^+ q(0) \right]\,  
 \left[ \bar{u}(\tilde{k},\lambda) \, H
           v(\tilde{k}',\lambda') \right]
 + {\cal O}\left( \frac{\Lambda^2}{s} \right) \,,
\hspace{2em}
\label{field-op}
\end{eqnarray}
where the restriction $\lambda=-\lambda'$ implements that the hard
scattering conserves chirality to leading order in the current quark
masses $m_q$.  The product of two bad components in
$\ov{q}{}_\alpha(x)\, q_\beta(0)$ is suppressed by $\Lambda^2/s$.
Terms with one good and one bad component are even smaller: since they
flip quark chirality in the hard scattering they come with a factor of
$m_q/\sqrt{s}$ in addition to the $\Lambda/\sqrt{s}$ suppression from
the soft matrix element.  With a suitable phase convention for quark
spinors (see e.g.~\cite{die2001}) and with antiquark spinors
satisfying $v(k,\lambda) = -u(k,-\lambda)$ we have
\begin{equation}
u(\tilde{k},\lambda)\, \ov{v}(\tilde{k'},-\lambda) = 
- \frac{1}{\sqrt{4 k^+ k'^+}}\, \frac{1 + \lambda \gamma_5}{2} \,
  (\tilde{k} \cdot \gamma)\, \gamma^+ (\tilde{k'}\cdot \gamma)
\end{equation}
and get, up to corrections of order $\Lambda^2/s$,
\begin{eqnarray}
{\cal A}_{\mu\mu'} &=&
 - \sum_q (e e_q)^2  \int\! d^4 k \, \frac{1}{\sqrt{4 k^+ k'^+}}\,
\sum_{\lambda} \bar{u}(\tilde{k},\lambda) \, H_{\mu\mu'}(k,k')\,
           v(\tilde{k}',-\lambda) \nonumber \\
& & {}\times 
  \int\! \frac{d^4 x}{(2 \pi)^4} \, e^{-i\, k \cdot x} \,
          \langle \pi^+(p) \, \pi^-(p') \, | \, T \,
  \ov{q}(x) \, \gamma^+ \frac{1 + \lambda \gamma_5}{2}\, q(0) \, 
          |0 \rangle \,,
\label{handbag-2}
\end{eqnarray}
where we have made explicit the dependence on the photon helicities
$\mu$ and $\mu'$.  Let us now concentrate on the term with the vector
current $\ov{q}(x) \, \gamma^+ q(0)$ and come back to the axial
current term later.  {}From charge conjugation invariance we know that
the two pions produced in the two-photon collision are in a $C$ even
state, so that we can explicitly symmetrize their state vector in the
soft matrix element,
\begin{equation}
{\cal S} = \frac{1}{2}
\int\! \frac{d^4 x}{(2 \pi)^4} \, e^{-i\, k \cdot x} \,
  \Big\langle \frac{\pi^+(p) \, \pi^-(p') + \pi^+(p') \, \pi^-(p)}{2}\,
  \Big| \, T \,\ov{q}(x) \, \gamma^+  q(0) \, \Big|\, 0 \Big\rangle \,.
\label{soft-mat}
\end{equation}
Abbreviating
\begin{equation}
{\cal H}_{\mu\mu'} = \sum_\lambda \bar{u}(\tilde{k},\lambda) \,
        H_{\mu\mu'}(k,k')\, v(\tilde{k}',-\lambda)
\end{equation}
we then have
\begin{equation}
{\cal A}_{\mu\mu'} =
 - \sum_q (e e_q)^2  \int\! d^4 k \, \frac{1}{\sqrt{4 k^+ k'^+}}\,
 {\cal H}_{\mu\mu'}(k,k') \, {\cal S}(k,k')
+ \mbox{axial current term} \,,
\label{handbag-3}
\end{equation}
where due to charge conjugation invariance 
\begin{equation}
{\cal S}(k,k') = - {\cal S}(k',k)  \,, \qquad
{\cal H}(k,k') = - {\cal H}(k',k)  \,.
\label{odd}
\end{equation}
According to our hypothesis, the soft matrix element ${\cal S}(k,k')$
should be strongly peaked when~(\ref{near-collinear}) is fulfilled.
The two regions $k\approx p$ and $k\approx p'$ where this is the case
are related through a rotation by $\pi$ about the 3-axis of our
coordinate system.  To proceed we separate the integration over $k$
into two regions, one with $\varphi \in [-\pi/2,\, \pi/2]$ and one
with $\varphi \in [\pi/2,\, 3\pi/2]$.  Because of (\ref{odd}) both
give the same integral, and we can write
\begin{equation}
\int\! \frac{d^4 k}{\sqrt{4 k^+ k'^+}}\, 
  {\cal H}(k,k') \, {\cal S}(k,k')
= \int \frac{dk^+\, dk^-\, d \tr{k}^2}{\sqrt{4 k^+ k'^+}}
  \int_{-\pi/2}^{\pi/2}d\varphi\;
  {\cal H}(k,k') \, {\cal S}(k,k') \,,
\label{hard-soft}
\end{equation}
where the integral is dominated by the region $k\approx p$.  Since the
hard scattering ${\cal H}(k,k')$ depends significantly on $k$ and $k'$
only over scales of order $\sqrt{s}$, we Taylor expand it around
$z=1/2$, $\varphi=0$, $\delta^- = \delta_\perp = 0$.  Keeping
only the leading terms of the expansion in $\delta^-$ and
$\delta_\perp$ we obtain ${\cal H}_{--} = {\cal H}_{++} = 0$ and
\begin{equation}
{\cal H}_{+-} = {\cal H}_{-+} = 
 2\, \Big( \sqrt{ u/t } - \sqrt{ t/u } \, \Big)
 - (z-\bar{z}) \, \Big( s/t + s/u \Big) 
 + {\cal O}\Big( (z-\bar{z})^2, \varphi^2 \Big)  \,,
\label{taylor}
\end{equation}
where according to (\ref{near-collinear}) the first term is of order
1, the second of order $\Lambda^2/s$, and the terms denoted by ${\cal
O}$ of order $\Lambda^4/s^2$.  It turns out that the leading term in
the expansion (\ref{taylor}) leads to a zero integral in
(\ref{hard-soft}).  To see this we remark that due to rotation
invariance about the 3-axis we have ${\cal S}(k^+,k^-,\tr{k}) = {\cal
S}(k^+,k^-,-\tr{k})$, so that
\begin{equation}
  \int \frac{dk^+\, dk^-\, d \tr{k}^2}{\sqrt{4 k^+ k'^+}}
  \int_{-\pi/2}^{\pi/2}d\varphi\; {\cal S}(k,k') =
\int\! \frac{d^4 k}{\sqrt{4 k^+ k'^+}}\, {\cal S}(k,k')  \,,
\label{is-zero}
\end{equation}
which is zero because of (\ref{odd}).  The vanishing of what would
have been the leading term is thus due to a conspiracy of invariance
under charge conjugation and rotation, and it is instructive to see
why this does not happen in the crossed channel process, even if one
scatters on a $C$ eigenstate.  The soft handbag contribution to wide-angle 
Compton scattering $\gamma \pi^0\to \gamma \pi^0$ is given by a
convolution analogous to (\ref{handbag-3}).  The two possible
solutions to the condition that the hadronic matrix element is soft
now correspond to the photon scattering on a quark or on an antiquark.
The integration over the parton momentum $k$ is then split into
regions $k^+>0$ and $k^+<0$, and these two regions are only related by
charge conjugation, but \emph{not} by any rotation.

Due to parity invariance the axial current term in (\ref{handbag-3})
vanishes to leading order in the off-shell parameters $\delta^-$ and
$\delta_\perp$.  The first nonzero contribution to our process is then
the one going with $z-\bar{z}$ in (\ref{taylor}), which according to
(\ref{near-collinear}) and (\ref{field-op}) is parametrically of the
same order as the parton off-shellness effects and contributions from
the bad components of the fermion fields.  A treatment of those is
beyond the scope of this work and will among other things have to
address issues of gauge invariance.  Rather we will remain with the
good components and the on-shell approximation, where the
hard-scattering $\gamma\gamma\to q\bar{q}$ is manifestly gauge
invariant.  We must then at this stage consider our result as a model,
or a partial calculation of the soft handbag contribution.

To proceed we thus keep the $z-\bar{z}$ term in ${\cal H}$.  Since it
is $\varphi$ independent, the integral over $\tr{k}$ can be extended
to the full region as in (\ref{is-zero}).  For a given $k^+$ we then
perform the integrals over $\tr{k}$ and $k^-$.  They only concern the
soft matrix element ${\cal S}(k,k')$, and we obtain the two-pion
distribution amplitude in light-cone gauge~\cite{die1998a},
\begin{equation} 
\Phi_{2\pi}^q(z,\zeta=1/2,s) = 
  \int \frac{d x^-}{2 \pi} \, e^{-i\, z (p+p')^+\, x^-}
        \langle \pi^+(p) \, \pi^-(p')\, | \,\ov{q}(x) \, \gamma^+
        q(0)\, |0 \rangle_{x = [\,0,\, x^-,\tr{0}]}    \;.
\label{pipida}
\end{equation}
Here we have used that for $\zeta=1/2$ explicit symmetrization in the
pion momenta is not needed, and that the time-ordering of the quark
fields can be dropped after the $k^-$ integration \cite{die1998b}.  Up
to still higher orders we have $\sqrt{4k^+ k'^+} \approx 2 p^+$ in
(\ref{handbag-3}) and obtain our final result
\begin{equation}
 {\cal A_{+-}} = {\cal A_{-+}} = - 4 \pi \ale \, \frac{s^2}{t u}\,
                                                       R_{2\pi}(s)\,,
\label{final}
\end{equation}
where we have defined the annihilation form factor by
\begin{equation}
   R_{2\pi}(s) = \sum_q e_q^2 \, R^q_{2\pi}(s) \,,
\qquad
 R^q_{2\pi}(s) = \frac{1}{2} \int_0^1 d z \, (2\, z-1) \, 
        \Phi_{2\pi}^{q}(z,1/2,s) \,.
\label{moment}
\end{equation}
The operator corresponding to this form factor is the quark part of
the energy-momentum tensor.  This has positive $C$ parity, as needed
for a pion pair produced in two-photon annihilation.  Note that
integrating $\Phi^q_{2\pi}(z,\zeta,s)$ over $z$ without the weight
$(2z-1)$ leads to the form factor of the quark vector current, which
is $C$ odd.

The differential cross section of our process is given by
\begin{equation}
 \frac{d\sigma}{d t}(\gamma\gamma \to \pi^+\pi^-) = 
    \frac{8 \pi \ale^2}{s^2} \, \frac{1}{\sin^4\theta}\;
    \Big|R_{2\pi}(s)\Big|^{\,2} \,,
\label{dsdt-pipi}
\end{equation}   
and the cross section integrated over $\cos\theta$ from
$-\cos\theta_0$ to $\cos\theta_0$ reads
\begin{equation}
 \sigma(\gamma\gamma \to \pi^+\pi^-) = \frac{4 \pi \ale^2}{s}
      \left[ \frac{\cos\theta_0}{\sin^2\theta_0} 
      + \frac12 \ln\frac{1+\cos\theta_0}{1-\cos\theta_0} \right] \,
                \Big|R_{2\pi}(s)\Big|^{\,2}\,.
\label{sig}
\end{equation}
When comparing with experiment we will quote the integrated cross
section for $\cos\theta_0=0.6$,
\begin{equation}
\sigma(\gamma\gamma \to \pi^+\pi^-) = 425~{\rm nb\, GeV}^2 \,
                         \Big|R_{2\pi}(s)\Big|^{\,2} /s \,.
\label{eq:num}
\end{equation}

\vskip\baselineskip {\it Flavor symmetry.}  Our results (\ref{final})
to (\ref{dsdt-pipi}) easily generalize to the production of other
pairs of pseudoscalar mesons.  New form factors then appear, which are
related by flavor symmetry.  A characteristic feature of the handbag
approach is the intermediate $q\bar{q}$ state, which allows only for
isospin $I=1$ and $I=0$.  Since a $\pi^+\pi^-$ pair in an $I=1$ state
is $C$-odd, $\pi^+\pi^-$ as well as $\pi^0\pi^0$ pairs are only
produced in isospin zero states.  This leads to \cite{die2000}
\begin{equation}
 R^u_{2\pi}(s) = R^d_{2\pi}(s)\,,
\end{equation} 
and to the same form factors for both charge combinations, resulting
in\footnote{The sign in this and the following relations depends on
the phase convention for the different meson states, cf.\ Appendix~A
of~\protect\cite{die2000}.}
\begin{equation}
  {\cal A}_{\mu\mu'}(\gamma\gamma \to \pi^+\pi^-) = 
    {\cal A}_{\mu\mu'}(\gamma\gamma \to \pi^0\pi^0) \,. 
\label{pi0-amp}
\end{equation}

Taking recourse to $U$-spin symmetry, i.e., the symmetry under the
exchange $d \leftrightarrow s$, we can relate the form factor for the
production of a $K^+K^-$ pair to that for pion pairs.  Since the
photon behaves as a $U$-spin singlet while $(K^+, \pi^+)$ and $(K^-,
\pi^-)$ are doublets, $U$-spin conservation leads to
\begin{equation}
\rule[-0.5em]{0pt}{2em}
  {\cal A}_{\mu\mu'}(\gamma\gamma \to K^+ K^-) \simeq 
    {\cal A}_{\mu\mu'}(\gamma\gamma \to \pi^+\pi^-)
\label{k-amp} 
\end{equation}
and corresponding relations among the two sets of form factors.  In
contrast to (\ref{pi0-amp}), which is characteristic of the handbag
approach, (\ref{k-amp}) holds in any dynamical approach respecting
SU(3) flavor symmetry.  Finally, isospin links the $K^+K^-$ form
factors $R^q_{2K}(s)$ to those for $K^0 \overline{K}{}^0$
production. Putting everything together we have the following set of
relations:
\begin{eqnarray}
R^u_{2\pi}(s) &=& R^u_{2\pi^0}(s) \;\simeq\; R^u_{2K}(s)
                              \;=\; R^d_{K^0\overline{K}{}^0}(s) 
\nonumber\\
              &=& R^d_{2\pi^0}(s) \;\simeq\; R^s_{2K}(s)
                              \;=\; R^s_{K^0\overline{K}{}^0}(s) \,,
\label{val} \\
R^s_{2\pi}(s) &=& R^s_{2\pi^0}(s) \;\simeq\; R^d_{2K}(s)
                              \;=\; R^u_{K^0\overline{K}{}^0}(s) \,.
\label{non-val}
\end{eqnarray}
We neglect isospin breaking while, in general, flavor symmetry
violations cannot numerically be ignored. This is indicated in
(\ref{k-amp}) to (\ref{non-val}) by the approximate symbol.

To the extent that flavor symmetry holds there are only two
independent form factors, a valence quark one, $R_{2\pi}^u$, and a
non-valence one, $R_{2\pi}^s$.  Following our discussion of the soft
handbag amplitude one may expect that $|R_{2\pi}^s| \ll |R_{2\pi}^u|$.
In order to be soft these form factors require the parton entering the
meson to take most of its momentum, and it is plausible to assume that
this parton is most likely a valence quark.  This is in accordance
with experience from parton densities in the limit $x\to 1$.  Except
for $K^0\overline{K}{}^0$ production, the contribution from the
non-valence form factor to the amplitudes is further suppressed by the
charge factor $e_d^2/(e_u^2+e_d^2)=1/5$.  For pions it thus seems to
be quite safe to neglect the non-valence form factor and we arrive at
\begin{equation}
\rule[-0.5em]{0pt}{2em}
        R_{2\pi}(s) = (e_u^2 + e_d^2)\, R_{2\pi}^u(s)\,.
\label{val-dom-pi}
\end{equation}
For the process $\gamma\gamma \to K^0\overline{K}{}^0$, on the other
hand, we have
\begin{equation}
  R_{K^0\overline{K}{}^0} (s) \simeq 
    e_u^2\, R^d_{2K}(s) + (e_d^2 + e_s^2)\, R^u_{2K}(s)\,,
\end{equation}
and see that this process is more sensitive to the non-valence form
factor.  Neglecting the non-valence contribution nevertheless, we
obtain
\begin{equation}
 R_{K^0\overline{K}{}^0} (s) \simeq 
    \frac{e_d^2 + e_s^2}{e_u^2 + e_s^2}\, R_{2K}(s)
\label{val-dom-K}
\end{equation}
and a corresponding relation between ${\cal A}_{\mu\mu'}(\gamma\gamma
\to K^+ K^-)$ and ${\cal A}_{\mu\mu'}(\gamma\gamma \to
K^0\overline{K}{}^0)$.

Notice that all we have used in our discussion of flavor symmetry is
the general structure of the handbag amplitude~(\ref{handbag-amp})
with its $q\bar{q}$ intermediate state, plus valence quark dominance
in the case of (\ref{val-dom-pi}) and (\ref{val-dom-K}).  Our
predictions relating $\pi^+\pi^-$ with $\pi^0\pi^0$ and $K^+ K^-$ with
$K^0\overline{K}{}^0$ production do therefore not require the
technical approximations we needed to arrive at (\ref{final}) and
(\ref{moment}), like the neglect of off-shell corrections or of the
bad components of the quark fields.  As already mentioned, the
relation (\ref{k-amp}) is yet more general.

\vskip\baselineskip {\it Comparison with experiment.}  The new
measurements of $\gamma \gamma\to \pi^+\pi^-,\;K^+K^-$ performed by
ALEPH \cite{aleph} and DELPHI \cite{delphi} allow for an experimental
determination of the annihilation form factors quite analogous to the
measurements of electromagnetic form factors.  One thus extracts
moments of the two-pion distribution amplitude.  This amplitude is
related by crossing to the ordinary parton distributions in the pion,
which have been extracted from Drell-Yan data in pion-nucleon
scattering.  The annihilation form factors and the two-pion
distribution amplitude can as yet not be calculated within QCD.  As
follows from our earlier remark they do not admit a direct
representation as overlaps of light-cone wave functions \cite{DFJK3}
either.  A recent investigation \cite{tib2001} has sought to
circumvent this restriction using a Bethe-Salpeter approach.  To our
knowledge, no model calculation is presently available for the
annihilation form factors in the $s$ range where we need them.

\begin{figure}[ht]
\begin{center}
\psfig{file=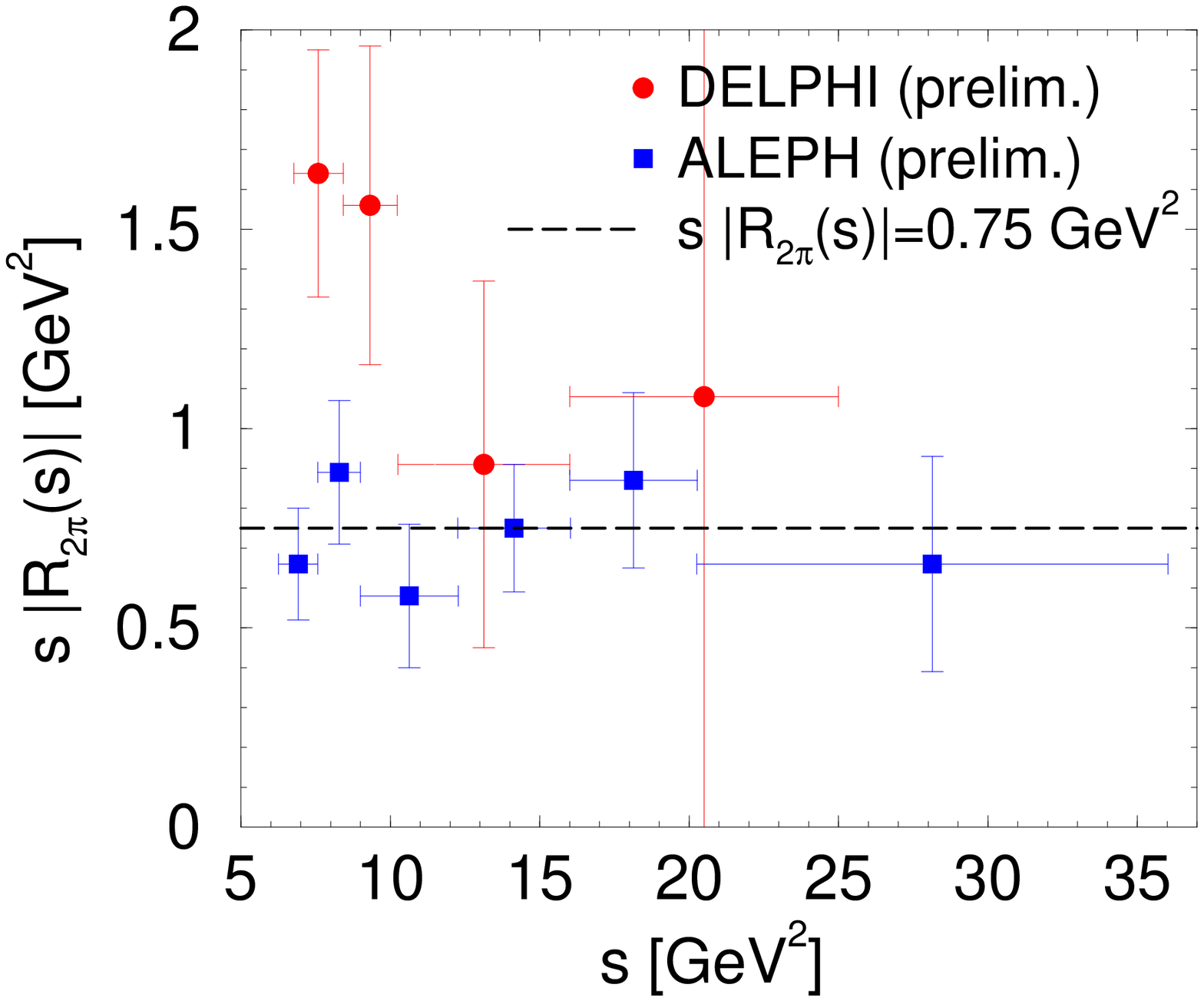,width=8cm}
\psfig{file=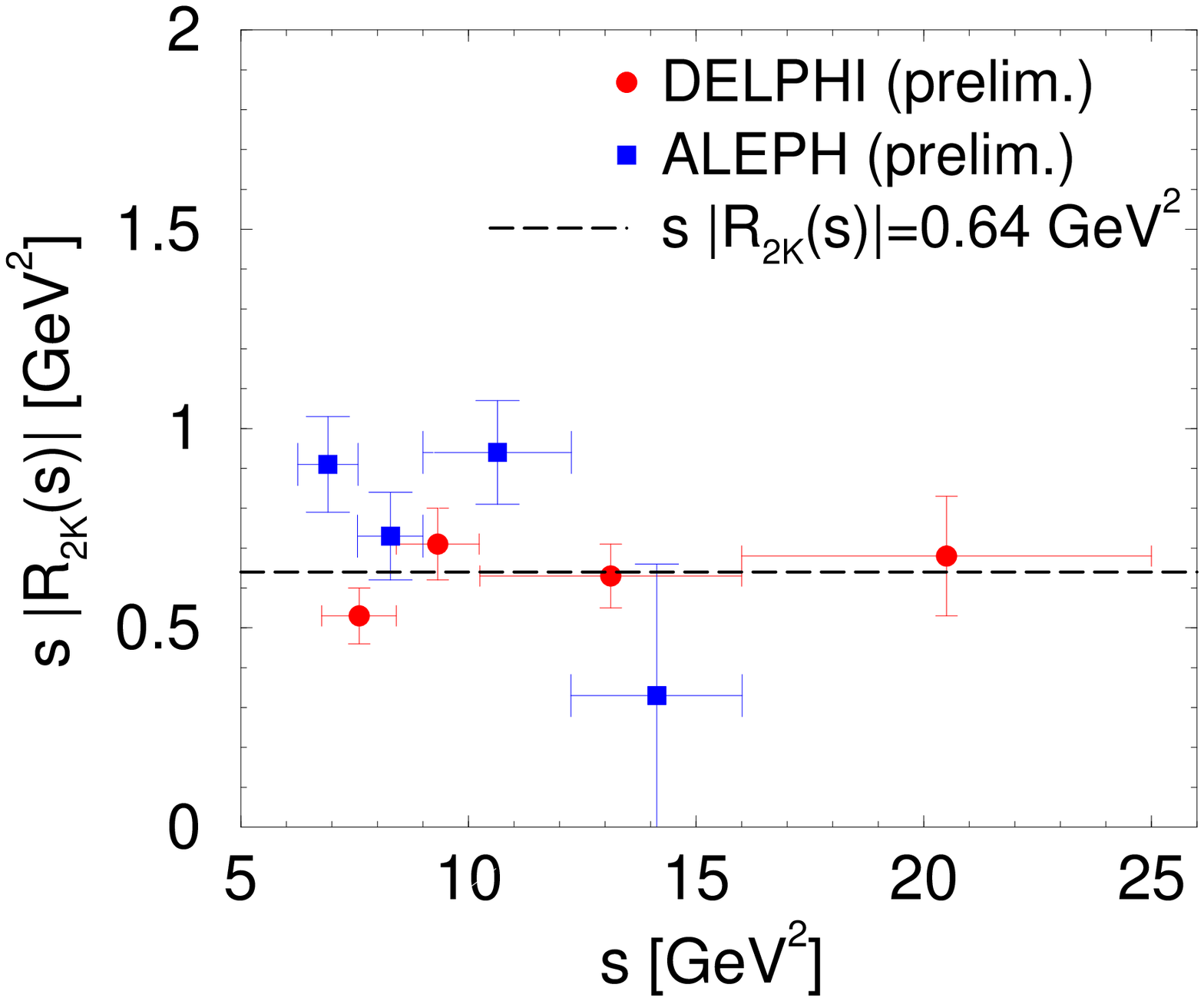,width=8cm}
\caption{The scaled annihilation form factors $s|R_{2\pi}|$ (left) and
$s|R_{2K}|$ (right) versus~$s$. The preliminary ALEPH and DELPHI data
is taken from \protect\cite{aleph,delphi} and plotted according to
(\protect\ref{eq:num}).  Dashed lines represent our fitted values
(\protect\ref{eq:num-ff}) and (\protect\ref{eq:fit-kaon}).}
\label{fig:ff}
\end{center}
\end{figure}

In order to avoid the resonance region, we restrict ourselves to data
with $\sqrt{s} \gsim 2.5 \gev$ here and in the following.  We have
used (\ref{eq:num}) to extract the form factor $R_{2\pi}(s)$ from the
preliminary data on $\gamma\gamma\to \pi^+\pi^-$ \cite{aleph,delphi}.
As Fig.~\ref{fig:ff} reveals, the form factor scaled by $s$ is
compatible with a constant over a large range of~$s$, within the still
large experimental errors.  A fit provides
\begin{equation}
              s |R_{2\pi}(s)| = 0.75 \pm 0.07\gev^2\,.
\label{eq:num-ff}
\end{equation}
The annihilation form factor is comparable in magnitude with the
timelike electromagnetic form factor of the pion, which is related to
the first moment of the two-pion distribution amplitude
\cite{die1998a,pol1998}.  We have performed a combined fit to the
admittedly poor $e^+e^-\to \pi^+\pi^-$ data \cite{bollini} in the
range $4 \gev^2 < s < 9 \gev^2$, and to the branching ratio of
$J/\Psi\to \pi^+\pi^-$ \cite{PDG}, which to a good approximation
provides the form factor at $s=M^2_{J/\Psi}$ \cite{bro1981}. This
yields
\begin{equation}
              s |F_{\pi}(s)| = 0.93 \pm 0.12 \gev^2\,.
\label{eq:elm-ff}
\end{equation}
Omitting the $J/\Psi$ data would increase the errors but not
significantly alter the central value of the fit.  It is amusing to
note that the data giving access to the annihilation form factor is
more precise than the one for the well-known and extensively discussed
electromagnetic one, where improvement would be highly welcome.  The
similarity between (\ref{eq:num-ff}) and (\ref{eq:elm-ff}) is
reminiscent of the spacelike region, where the form factors for
wide-angle Compton scattering off protons and the Dirac form factor
also have similar $s$ behavior and are of comparable magnitude
\cite{DFJK1}.

The $s$ dependence of both the annihilation and the electromagnetic
form factor is in agreement with the dimensional counting rule
behavior.  At this point we must realize that, since we have
calculated the soft part of the handbag diagrams, the form factor
appearing in our result (\ref{final}) is only the soft part $R^{\rm \,
soft}_{2\pi}(s)$ of the matrix element defined by (\ref{pipida}) and
(\ref{moment}).  At very large~$s$ this is power suppressed compared
to the hard perturbative part $R_{2\pi}^{\rm \, pert}(s)$, which
scales like $1/s$.  Asymptotically the cross section (\ref{dsdt-pipi})
for $\gamma\gamma\to \pi\pi$ therefore decreases faster than $1/s^{4}$
at fixed angle $\theta$ and thus is indeed a power correction to the
leading twist contribution.  Our fit (\ref{eq:num-ff}) of the form
factor to the available data does not display a falloff faster than
$1/s$.  In the absence of a dynamical model for $R^{\rm \,
soft}_{2\pi}(s)$ we cannot say at which $s$ this falloff will start
and how rapid it will be.  In the case of wide-angle Compton
scattering on the proton, an explicit model in terms of light-cone
wave functions has shown how the soft overlap part of the Compton form
factor can mimic dimensional counting behavior over a finite range of
$s$ \cite{DFJK2}.

As to the hard part $R_{2\pi}^{\rm \, pert}$ of the annihilation form
factor, it is readily obtained from the leading-twist expression of
the two-pion distribution amplitude at large $s$ \cite{DFKV}.  Taking
the asymptotic form of the single-pion distribution amplitude we get
$s|R_{2\pi}^{\rm \, pert}|\simeq \alpha_s \times 0.1 \gev^2$.  This
is indeed negligible compared to~(\ref{eq:num-ff}), and for simplicity
we write $R_{2\pi}(s)$ instead of $R^{\rm \, soft}_{2\pi}(s)$
throughout this work.

Clearly, the handbag diagrams are not the only ones to provide a soft
physics contribution to $\gamma\gamma\to \pi\pi$.  A different
contribution coming to mind is due to the hadronic components of the
photons, which one may model using vector meson dominance.
Unfortunately, no data is available for elastic or quasielastic
meson-meson scattering at large c.m.\ energy and angle.  We can only
observe that experimentally many other hadronic processes at large
angle show an $s$ behavior compatible with dimensional scaling.  If
this were also true for $\rho\rho\to \pi\pi$ then the vector dominance
contribution to two-photon annihilation would decrease faster than the
data in Fig.~\ref{fig:ff} by a power of $1/s$ at the amplitude level.
We also remark that for wide-angle Compton scattering off the proton
one can estimate the vector dominance part if, following quark model
ideas, one relates $\rho p\to \rho p$ to $\pi p\to \pi p$.  Using the
data for the latter, one finds that for $\theta \approx 90^\circ$ and
$s$ between 8 and $10 \gev^2$ the corresponding contribution to
$\gamma p\to \gamma p$ is about an order of magnitude below the
measured cross section \cite{Huang:2000kd}.  One also observes that
its suppression scales like $1/s$ in the amplitude.  We finally remark
that the flavor symmetry relations we elaborated for the soft handbag
are not generically satisfied by the vector dominance mechanism, since
a $\rho^0\rho^0$ pair can couple to isospin $I=2$.

In Ref.~\cite{pol1998} a simultaneous expansion of the two-pion
distribution amplitude in eigenfunctions of the corresponding
evolution kernel and in partial waves of the $\pi\pi$ system has been
given. The moment (\ref{moment}) of the two-pion distribution
amplitude involves two of the coefficients in that expansion:
\begin{equation}
R_{2\pi}(s)
   = \frac{5}{18} \int_0^1 dz (2z-1) \Phi^u_{2\pi}(z,1/2,s)
   = \frac{1}{6} \left[ B^u_{10}(s) - \frac12 B^u_{12}(s) \right] \,, 
\end{equation}
where we have neglected the non-valence contribution $R_{2\pi}^s$.
Here $B^u_{nl}$ is the expansion coefficient of the two-pion
distribution amplitude for $u$ quarks, with $n$ giving the order of
the Gegenbauer polynomial, and $l$ the partial wave of the $\pi\pi$
system.  We remark that for $s=0$ the coefficient $B_{12}$ can be
expressed in terms of the ratio $M_Q$ of momentum carried by quarks in
a single pion \cite{die2000,pol1998}. Furthermore, a soft pion theorem
\cite{pol1998} provides the relation $B_{10}(0) = - B_{12}(0)$.  With
these two inputs we obtain
\begin{equation}
\Big| R_{2\pi}(s=0) \Big| = \frac{5}{36}\, M_Q \,.
\label{fraction}
\end{equation}
Taking the LO GRS parameterization of parton distributions in the pion
\cite{reya} one finds $M_Q$ between 0.7 and 0.5 at renormalization
scales $\mu^2$ from $0.26\gev^2$ to $36\gev^2$.  
The size of $R_{2\pi}(s)$ at $s=0$ is thus comparable to the one which 
our fit~(\ref{eq:num-ff}) gives for $s$ around $6 \gev^2$, 
just above the resonance region.

\begin{figure}[ht]
\begin{center}
\psfig{file=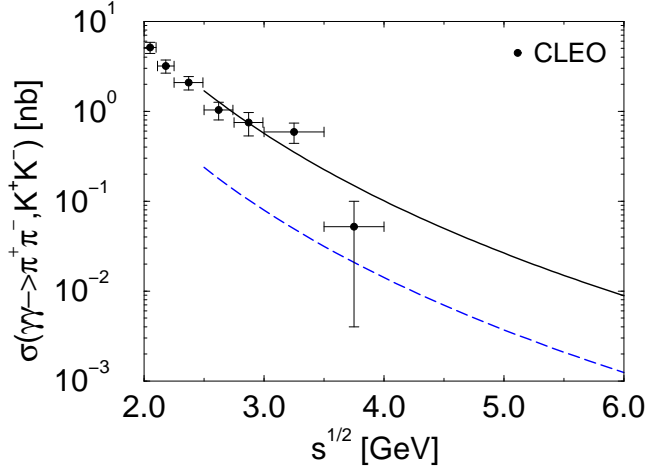,width=7.8cm}
\caption{The CLEO data \protect\cite{cleo} for the cross section
$\sigma(\gamma\gamma\to \pi^+\pi^-) + \sigma(\gamma\gamma\to K^+K^-)$
integrated with $|\cos{\theta}| < 0.6$.  The solid line is the result
of the handbag approach with our fitted annihilation form factors
(\protect\ref{eq:num-ff}) and (\protect\ref{eq:fit-kaon}).  The dashed
line is the estimate of the leading-twist perturbative contribution
described below.}
\label{fig:cleo}
\end{center}
\end{figure}

The analysis of the preliminary data for the production of charged
kaon pairs, see Fig.~\ref{fig:ff}, gives for the kaon annihilation
form factor
\begin{equation}
s |R_{2K}(s)| = 0.64 \pm   0.04 \gev^2 \,,
\label{eq:fit-kaon}
\end{equation}
which is close to the value (\ref{eq:num-ff}) for charged pions.
Taking the central values of our fits we find that flavor symmetry
violation lead to a suppression of the kaon form factor by about
$15\%$, which according to phenomenological experience is a rather
typical value.  For the cross sections our fits (\ref{eq:num-ff}) and
(\ref{eq:fit-kaon}) give a ratio of $K^+K^-$ to $\pi^+\pi^-$
production between 0.54 and 1 within one standard deviation, with a
central value of 0.73.  We regard this as compatible with the $U$-spin
relation
\begin{equation}
  \frac{d\sigma}{dt}(\gamma\gamma \to K^+ K^-) \simeq 
        \, \frac{d\sigma}{dt}(\gamma\gamma \to \pi^+\pi^-) \,. 
\end{equation}
In Fig.~\ref{fig:cleo} we compare our results with the CLEO data for
the integrated cross section \cite{cleo}, where pions and kaons have
not been separated, and find rather good agreement.

\begin{figure}[ht]
\begin{center}
\psfig{file=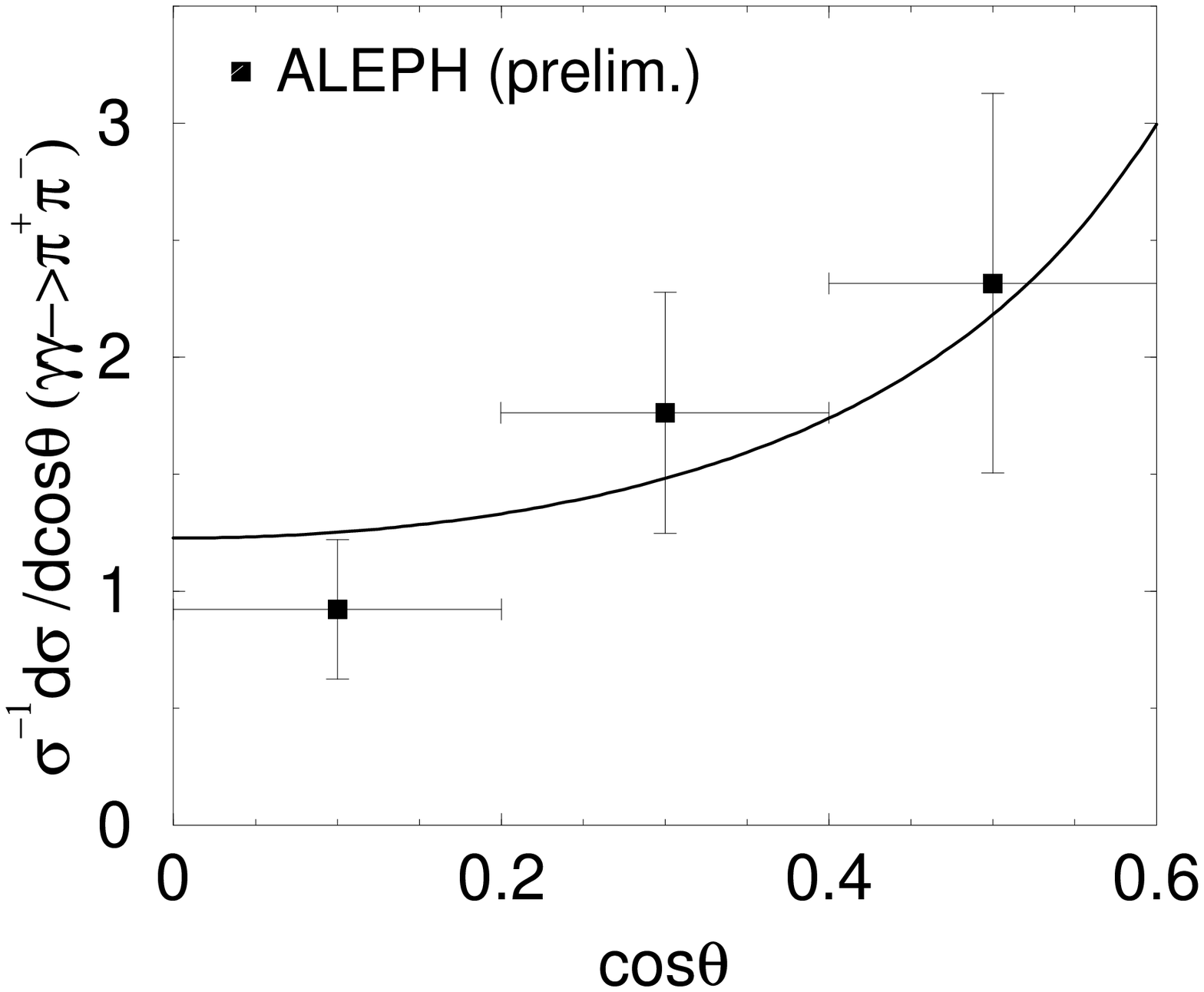,width=7.3cm}
\psfig{file=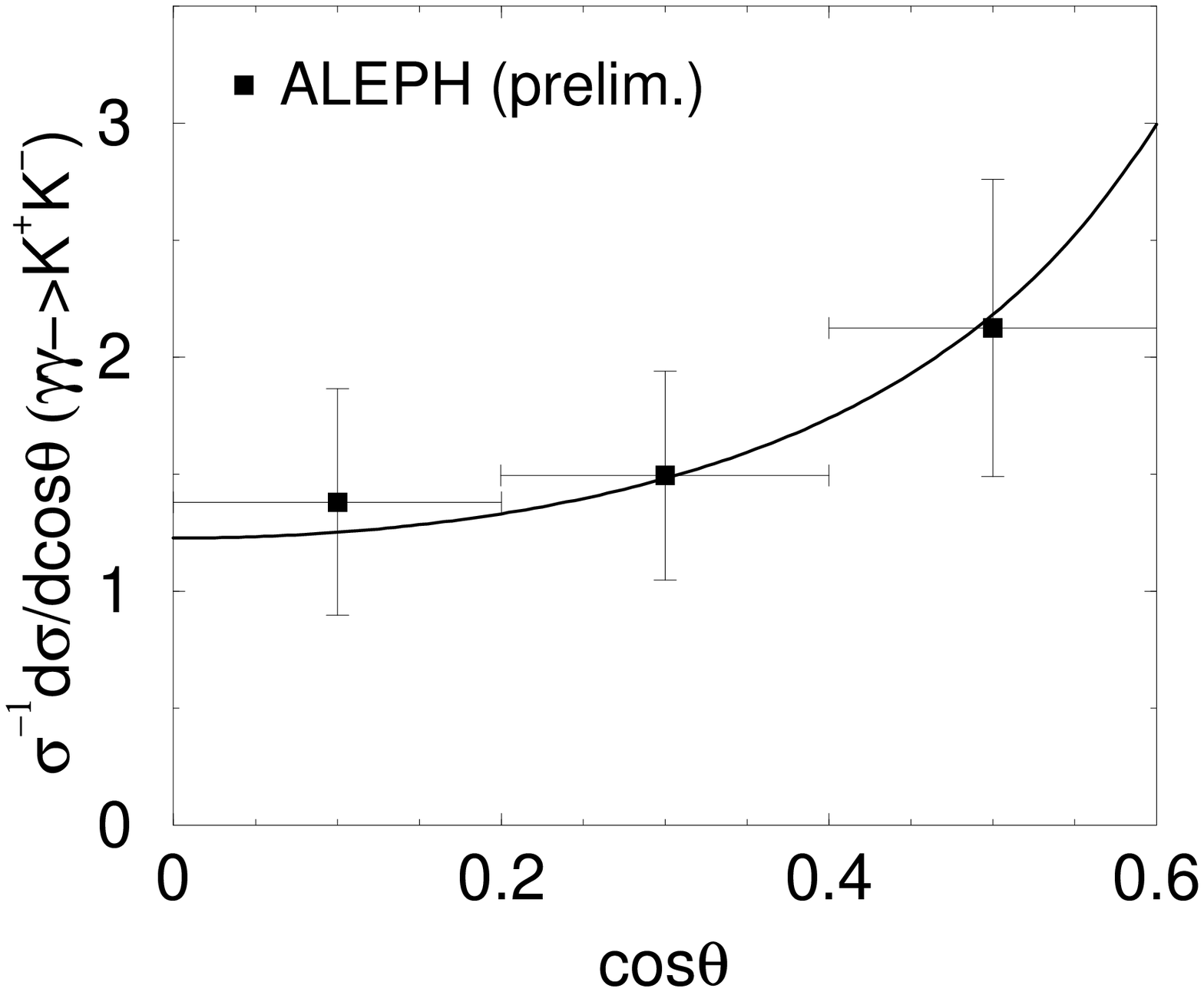,width=7.3cm}
\caption{The result of the handbag calculation for the angular
dependence of the cross section for $\gamma\gamma\to\pi^+\pi^-$ (left)
and $\gamma\gamma\to K^+K^-$ (right), compared to the preliminary
ALEPH data for $4 \gev^2 <s <36 \gev^2$ \protect\cite{aleph}.  We have
normalized both data and theory to give unit area under the curve.}
\label{fig:ang-dis}
\end{center}
\end{figure}

The $1/\sin^4 \theta$ behavior of the differential cross section,
which represents a characteristic result of our handbag calculation
(\ref{dsdt-pipi}), is confronted with experiment in
Fig.~\ref{fig:ang-dis}.  Good agreement with the preliminary ALEPH
data \cite{aleph} for pions and kaons can be observed. The large $s$
data from the other experiments \cite{delphi,cleo} are comparable with
a $1/\sin^4 \theta$ behavior, too.

We recall that, besides the angular dependence, there is another
parameter-free prediction of the handbag approach:
\begin{equation}
  \frac{d\sigma}{dt}(\gamma\gamma \to \pi^0\pi^0) = 
       \, \frac{d\sigma}{dt}(\gamma\gamma \to \pi^+\pi^-) \,. 
\end{equation}
In the integrated cross section the statistical factor $1/2$ for
identical particles in the final state must be taken into account.
Unfortunately the existing data for the $\pi^0\pi^0$ channel is either
at too small energies or has too large errors.  Likewise, no data is
available on $K^0\overline{K}{}^0$ production at $\sqrt{s} \gsim
2.5~\gev$.  To the extent that the non-valence form factor can be
neglected we find
\begin{equation}
\frac{d\sigma}{dt}(\gamma\gamma\to K^0\overline{K}{}^0) 
  \simeq \frac{4}{25} \,
                  \frac{d\sigma}{dt}(\gamma\gamma \to K^+K^-)\,.
\label{oo}
\end{equation}
It would be interesting to examine this relation experimentally.
Deviations from (\ref{oo}) may provide information on the non-valence
form factor $R^d_{2K}$ and on the pattern of flavor SU(3) breaking.
The generalization to other pseudoscalar pairs like $\eta\,\eta$ and
$\eta'\eta'$ is also possible.  Based on flavor symmetry we expect
production rates of similar size as for $K^0\overline{K}{}^0$.

\vskip\baselineskip {\it Comparison with the leading-twist
perturbative approach.}  Let us now discuss a few characteristic
differences between our handbag approach and the hard scattering
picture of Brodsky and Lepage.  In the perturbative approach there are
two $q \bar{q}$ pairs in the intermediate state, which allows for a
non-zero isospin $I=2$ amplitude.  Hence, our relation~(\ref{pi0-amp})
does not hold in this approach.  In fact, if one uses a pion
distribution amplitude which is compatible with the photon-pion
transition form factor~\cite{kro96,bro98,DKV2001}, the differential
cross section for $\pi^0 \pi^0$ production is found about an order of
magnitude smaller than that for $\pi^+ \pi^-$ pairs~\cite{bl1981}.
This implies $I=0$ and $I=2$ transitions of nearly the same magnitude,
in sharp contrast to the situation in the handbag approach.  Flavor
SU(3) violations also manifest themselves differently in the two
mechanisms.  Since the single-meson distribution amplitudes are
normalized to the respective decay constants, a factor of
$(f_K/f_\pi)^4 \approx 2.2$ appears in favor of the $\gamma\gamma \to
K^+ K^-$ cross section in the perturbative picture.  In order to
obtain a $K^+K^-$ cross section comparable to or smaller than the one
for $\pi^+\pi^-$, one needs a narrower shape for the kaon than for the
pion distribution amplitude.

As already mentioned in the introduction, the perturbative result is
way below the experimental data if single-pion distribution amplitudes
consistent with other data are employed~\cite{vogt2000}.  Studies of
the spacelike pion form factor $F_{\pi}(Q^2)$ suggest that for
processes of the type we are considering higher order corrections in
$\alpha_s$ can be substantial.  One may hope to keep their size
moderate by using the BLM prescription for setting the scale of the
running coupling~\cite{Brodsky:1983gc}.  This cannot be done for our
process as long as the next-to-leading order corrections in $\alpha_s$
have not been calculated, but one may take the spacelike pion form
factor as a guideline, where $\mu^2_{\it BLM} \approx 0.05\, Q^2$
\cite{Melic:1999qr}.  For most of the $s$ range we are dealing with
the corresponding scale is then too low to use the perturbative
expression of the running coupling, and one has to make an ansatz for
its behavior in the infrared region.  This is a highly nontrivial
problem, and a wide choice of options is discussed in the literature.
For simplicity we evaluate the leading-twist expression with a fixed
coupling $\alpha_s=0.5$, a size suggested by different lines of
investigation~\cite{Dokshitzer:2001yu}.  Following~\cite{vogt2000} we
take the asymptotic form for both pion and kaon distribution
amplitudes, which in light of our above remark should rather over-
than underestimate the $K^+K^-$ cross section.  The leading-twist
prediction thus obtained amounts to about 15\% of our fitted handbag
result as shown in Fig.~\ref{fig:cleo}.  In view of this we consider
that we make an acceptably small error in our analysis by altogether
neglecting the hard perturbative part compared with the soft handbag
mechanism.  Note that taking it into account would require us to fit
both the magnitude and the phase of $R_{2\pi}(s)$ since the two
contributions must be added at amplitude level.  Also, a careful study
would be necessary to avoid double counting because the leading-twist
expression evaluated with an infrared saturated coupling contains soft
physics effects, including the diagrams with handbag topology.

Brodsky and Lepage~\cite{bl1981} have proposed a formula for meson
pair production which looks similar to~(\ref{dsdt-pipi}), except for a
different charge factor and the appearance of the timelike
electromagnetic meson form factor instead of the annihilation form
factor $R(s)$.  This formula was obtained from the leading-twist
result by neglecting part of the amplitudes with opposite photon
helicities.  As has been pointed out in~\cite{ben1989}, this part is
however not approximately independent of the pion distribution
amplitude and not generically small.  We also remark that the
appearance of $F_\pi(s)$ in the $\gamma\gamma\to \pi^+\pi^-$ amplitude
is no longer observed if corrections from partonic transverse momentum
in the hard scattering process are taken into account, and that these
corrections are not numerically small for the values of $s$ we are
dealing with~\cite{vogt2000}.  Notice further that two-photon
annihilation produces two pions in a $C$-even state, whereas the
electromagnetic form factor projects on the $C$-odd state of a pion
pair.  In contrast, our annihilation form factor $R_{2\pi}(s)$ is
$C$-even as discussed after~(\ref{moment}).  Finally, due to a
particular charge factor, the Brodsky-Lepage formula leads to a
vanishing cross section for $\gamma\gamma$ annihilation into pairs of
neutral pseudoscalars.

On the other hand, its apparent phenomenological success for
$\pi^+\pi^-$ and $K^+ K^-$ production is not a surprise because of its
similarity to our result~(\ref{dsdt-pipi}).  This success is achieved
if one takes a suitable value for the timelike electromagnetic form
factor, related to our annihilation form factor via $|F_\pi(s)|^{\it
BL} = |R_{2\pi}(s)|/\sqrt{2}$.  Our fit (\ref{eq:num-ff}) amounts to
$s|F_\pi|^{\it BL} = 0.53 \pm 0.05 \gev^2$, which is clearly smaller
than the experimental value (\ref{eq:elm-ff}) we extracted for
$s|F_\pi|$, and at the same time larger than the leading-twist
perturbative result for this form factor given in \cite{bro98}.  In
view of this we do not think that the presently available data on
$F_\pi(s)$ and on $\gamma\gamma$ annihilation into $\pi^+\pi^-$ and
$K^+K^-$ can be considered as a success of the Brodsky-Lepage formula
or of the leading-twist perturbative approach.

\vskip\baselineskip {\it Summary.}  We have discussed the soft handbag
contribution to two-photon annihilation into pseudoscalar meson pairs
at large energy and large momentum transfer.  Our main result is to
express the amplitude as a product of a parton-level subprocess,
$\gamma\gamma \to q\bar{q}$, and an annihilation form factor given by
a moment of the two-meson distribution amplitude at skewness
$\zeta=1/2$.  The operator associated with this form factor is the
quark part of the energy-momentum tensor.  To obtain our result we
have neglected quark off-shell effects in the hard scattering and the
bad components of the corresponding field operators.  A closer
investigation of these corrections, which as far as we could establish
are of the same parametric order as the terms we retained, is an open
task.  We remark that according to Radyushkin it may be possible to
treat the processes under investigation in the framework of double
distributions \cite{rad:bad}.

Although the handbag contribution formally represents a power
correction to the asymptotically leading perturbative contribution, it
seems to dominate at experimentally accessible energies.  We find that
the data for $\pi^+\pi^-$ and $K^+K^-$ production is compatible with
annihilation form factors behaving as $1/s$ for $s$ between $6\gev^2$
and $36\gev^2$, a counting rule behavior typical of many exclusive
observables.  Fitting the form factors to the preliminary
ALEPH~\cite{aleph} and DELPHI~\cite{delphi} data, we find that for
pions the annihilation form factor is comparable in size to the
timelike electromagnetic form factor, and that for kaons it is
suppressed by an amount consistent with moderate flavor SU(3)
breaking.  A severe test of our approach is the $1/\sin^4 \theta$
angular dependence of the cross section, which agrees well with the
preliminary ALEPH and DELPHI data.  We also find good agreement with
the CLEO data~\cite{cleo} on the combined cross section for pion and
kaon production.

A key prediction of the handbag mechanism is that the differential
cross sections for $\pi^+\pi^-$ and $\pi^0\pi^0$ production should be
the same. This is in sharp contrast to the leading-twist perturbative
approach, where $\pi^0\pi^0$ is found suppressed by about an order of
magnitude.  Measurement of the production ratio of neutral and charged
pion pairs would thus be most valuable to help us understand the
dynamics of such processes.  Under further assumptions, the handbag
mechanism also predicts the production ratio of $K^0\overline{K}{}^0$
and $K^+K^-$.

Amusingly, our expression for the cross section is very similar to the
formula proposed by Brodsky and Lepage~\cite{bl1981}, where instead of
our annihilation form factor the timelike electromagnetic one appears.
We would however like to emphasize that the dynamical origins of the
two expressions are completely different.  We also recall that the
Brodsky-Lepage formula does not represent the full leading-twist
perturbative result, and we found that it has normalization problems
when compared with presently available data.

The factorization of the soft handbag diagrams is analogous to the one
in wide-angle Compton scattering.  For the latter it has recently been
shown that this factorization remains valid when taking into account
next-to-leading corrections in $\als$ to the parton-level subprocess
\cite{hkm}.  We are tempted to expect that this also holds in the
annihilation process considered here.

It is straightforward to extend the results of this letter to the case
where one or two of the photons is off-shell by an amount not
significantly bigger than the large scale $s$ in the process.  Another
generalization is the production of vector meson or baryon-antibaryon
pairs, where several form factors describing the spin structure of the
final state will appear.  Finally, the time reversed process of
$p\bar{p}$ annihilation into photon pairs can be described in the same
way.

\vskip\baselineskip {\it Acknowledgments.}  We would like to thank
E.~Leader for correspondence, and A.~Finch and K.~Grzelak for
providing us with the numbers of the preliminary ALEPH and DELPHI
data.


\end{document}